%% file: main.tex
\documentclass[10pt,journal,compsoc]{IEEEtran}

\usepackage{booktabs,ragged2e}         
\usepackage[flushleft]{threeparttable} 
\usepackage{multirow}                  
\usepackage{makecell}                  
\usepackage{graphicx}                  
\usepackage{pdfpages}                  
\usepackage{algorithm}                 
\usepackage{algpseudocode}             

\ifCLASSOPTIONcompsoc
  \usepackage[nocompress]{cite}
\else
  \usepackage{cite}
\fi
\ifCLASSINFOpdf
\else
\fi

\usepackage[nopatch=eqnum]{microtype}

\begin{document}


\title{Threat Trekker: An Approach to\\ Cyber Threat Hunting}

\author{\'Angel~Casanova~Bienzobas,~Alfonso~S\'anchez-Maci\'an~Perez
\IEEEcompsocitemizethanks{\IEEEcompsocthanksitem A. Casanova and A. Sanchez are with Universidad Carlos III de Madrid, Avda de la Universidad 30, Legan\'es, Madrid, Spain. Emails: 100487199\{at\}alumnos.uc3m.es, and alfonsan\{at\}it.uc3m.es. \\
}
}

\markboth{Paper, May~2023}%
{Threat Trekker \MakeLowercase{\textit{et al.}}: An Approach to Cyber Threat Hunting}

\IEEEtitleabstractindextext{%
\begin{abstract}
\input{sections/abstract}
\end{abstract}
}
\maketitle
\IEEEpeerreviewmaketitle

\IEEEraisesectionheading{\section{Introduction}\label{sec:introduction}}
\input{./sections/introduction}

\section{Related Works}\label{sec:related_works}
\input{./sections/related_works}

\section{Methodology}\label{sec:proposed_method}
\input{./sections/proposed_method}

\section{Results}\label{sec:results}
\input{./sections/results}

\section{Future Work}\label{sec:production}
\input{./sections/go_to_production}

\section{Conclusion}\label{sec:conclusion}
\input{./sections/conclusion}

\section*{Acknowledgments}

\input{sections/acknowledgments}
\ifCLASSOPTIONcaptionsoff
  \newpage
\fi

\bibliographystyle{bib/IEEEtran}
\bibliography{bib/main}

\end{document}

%% file: sections/abstract.tex
Threat hunting is a proactive methodology for exploring, detecting and mitigating cyberattacks within complex environments. As opposed to conventional detection systems, threat hunting strategies assume adversaries have infiltrated the system; as a result they proactively search out any unusual patterns or activities which might indicate intrusion attempts.
Historically, this endeavour has been pursued using three investigation methodologies: (1) Hypothesis-Driven Investigations; (2) Indicator of Compromise (IOC); and (3) High-level machine learning analysis-based approaches.
Therefore, this paper introduces a novel machine learning paradigm known as \textit{Threat Trekker}. This proposal utilizes connectors to feed data directly into an event streaming channel for processing by the algorithm and provide feedback back into its host network.
Conclusions drawn from these experiments clearly establish the efficacy of employing machine learning for classifying more subtle attacks.

%% file: sections/introduction.tex
\IEEEPARstart{C}{yber} Threat Hunting (CTH) is the practice of seeking and identifying cyber threats which have bypassed traditional security measures such as firewalls, antivirus software and intrusion detection systems \cite{survey1}.
Attaining this result typically involved training cybersecurity professionals on using specific tools and techniques to detect signs that indicate Advanced Persistence Threat (APT) within systems or networks \cite{9900201}.

Today's Internet's continued and exponential expansion makes this task impossible for humans alone to achieve due to all of the data that's produced today by modern networks \cite{10216378}; experts therefore develop machine learning models specifically for this task in order to complete it more quickly and successfully.
Given that CTH involves collecting network data such as traffic, logs or binaries from anywhere within a network environment, training the machine learning algorithm becomes extremely complicated due to normalizing and encoding data into something understandable for machine learning algorithms.

Contemporary scholarship has taken great strides to identify and address potential threats by employing an integrated three-pronged strategy: hypothesis-driven analysis frameworks; stringent inspection of indicators of compromise; and innovative machine learning approaches.

\subsection{Hypothesis-driven}\label{subsec:hypothesis-driven}
Hypothesis-driven investigations typically are carried out after an application or program has been identified as potentially undesirable by another third party, then from there collected and analysed to either confirm or disprove this assumption. The threat hunter then collects and examines data in order to either support or refute his initial assumption.
Process analysis includes continuously monitoring network and system logs, analyzing traffic patterns, and investigating anomalies or suspicious activities to detect possible security incidents.

Overall, hypothesis-driven investigation is an effective method for cyber threat hunting as it allows security professionals to focus their efforts on specific threats while prioritizing investigations based on potential impact of potential risks.

\subsection{Indicators~of~Compromise}\label{subsec:ioc}
Indicators of Compromise are pieces of information which suggest the presence of cyber threats or security incidents, including IP addresses, domain names, hashed files or patterns of behavior.
Threat hunters use indicators like these to recognize possible security threats, locate compromised systems and investigate security incidents.

Under this approach, threat hunters rely on automated tools, like Security Information and Event Management (SIEM), to collect and examine log data from both network systems and system logs.
Researchers then use this knowledge to search for known IOCs and use this information to form hypotheses about any possible threats that exist in their environment.

\subsection{Machine~Learning}\label{subsec:machine_learning}
Utilizing machine learning (ML) algorithms, threat hunters use this strategy to assess large volumes of network and system data and detect patterns or anomalies to spot possible security threats \cite{9845345}.
Over time, algorithms learn from data to become ever more precise in identifying potential threats.

One popular machine learning application used for threat hunting is anomaly detection algorithms. Such software learns what normal behavior should look like within an organization's network and systems and can identify deviations that indicate potential security threats.
As opposed to IOC approaches which rely solely on established indicators to detect threats, ML algorithms have the capacity to learn from data and identify previously unsuspected threats - making ML solutions more adaptive to new and evolving threats than IOC methods.

At its core, machine learning offers several advantages over IOC in cyber threat hunting: flexibility, efficiency, automation accuracy and continuous learning - which have seen its usage grow rapidly among modern cybersecurity strategies.

%% file: sections/related_works.tex
Recent years have witnessed an explosive expansion in internet surface area coupled with an inexorable rise of novel malware arising simultaneously with rapid deployment of innovative online services.
As malware analysis evolves in tandem, innovative methodologies have also flourished - most significantly through advances made in machine learning ~\cite{9574308} and \cite{10041568}. This paper investigates these advancements.

Noteworthy is the fact that such methodologies often adapt to particular contexts, suggesting there may be little standardization among various corporate entities for collecting and storing tracking information.

MITRE ATT\&K provides the industry \emph{de-facto} standard to identify tactics, technologies and procedures related to cyber threat hunting investigations \cite{alshaer2020learning}. It plays an increasingly influential role within CTH investigations by providing an efficient means for juxtaposing attack samples collected from disparate datasets into an efficient threat hunting approach.

Certain areas of inquiry center around uncovering botnets - an objective supported by research \cite{9169024}, and \cite{10205781}. This thematic area has garnered increased focus since notable botnet instances like Mirai \cite{8115504}, \cite{9685306}, and \cite{10019223} and Medussa \cite{7971869} have come to prominence and increased public understanding about them.

Analysis of traffic and logs generated by specific networks to implement threat hunting using machine learning models which meet their applications is also an innovative practice \cite{8661952}; its popularity can be attributed to cost effectiveness versus conventional solutions like building out and managing in-house Security Operations Center (SOC) teams.

Given their remarkable precision, innovative solutions like threat hunting have led to new research initiatives focused on exploiting its capabilities for finding zero-day vulnerabilities \cite{10041568}.

Properly aligning with its overarching purpose, the proposed algorithm seeks to leverage private networks' collected insights for uncovering any suspicious activities that might signal an APT threat. It does so through systematic searching methods designed to discover deviations indicative of unusual behavioral anomalies which might hint at such threats.

Attracting inspiration from recent advancements in threat hunting using machine learning, a strategic determination was reached to craft a distinctive approach of our own. This culminated in the creation of an original application known as \textit{Threat Trekker}.

As part of our evaluation strategy for our application, a deliberate choice was made to utilize various datasets in order to test its efficacy. By doing this, it allowed for validation that our proposition represented an adaptable solution suited for an array of network configurations and user scenarios. Training of the algorithm involved using \textit{UWF-2022} \cite{data8010018}, \textit{CIC-IDS2017} \cite{Sharafaldin2018TowardGA}, and \textit{TON\textunderscore IoT} \cite{9444348} datasets.; their creators took great care in collecting this research data themselves.

Though subsequent sections provide detailed analyses conducted with datasets, its purpose goes far beyond these experiments. Threat Trekker truly shines with its deployment alongside an ensemble of connectors designed to gather network data and feed it directly into an event streaming conduit for processing by machine learning algorithms - providing enhanced data fusion through log analysis combined with network metrics in an event streaming conduit. This fusion ultimately leads to an improve upon the level of accuracy.

Threat Trekker is reckoned to stand out in another respect with its future capability of disseminating training outcomes back onto a network through the main event streaming mechanism, making it ideal for integration with SIEM systems as well as defensive applications like antivirus. Through this integration alerts can be generated depending on attack types predicted allowing the integration of threat hunting within any broader network infrastructure.

Future Work section details Threat Trekker's future vision in great depth, yet to be implemented features remain unavailable due to the lack of an authentic environment suitable for conducting experiments. As this software manages information that includes Personally Identifiable Information (PII), corporations may hesitant to allow researchers accessing client's data as there may be legal ramifications associated with data protection laws and thus researchers cannot perform experiments using client information for experimentation purposes.

%% file: sections/proposed_method.tex
This project proposes the classification of log datasets instead of directly collecting information through network traffic sniffing, though authentic data would likely be preferred due to stringent data protection regulations which prohibit companies from sharing it due to concerns of potential data breaches.
Due to these constraints, gathering true functional network data proved impractical for this project. Therefore, the decision was made to acquire a dataset which provides up-to-date data resembling real life scenarios.

As part of our investigation, we tested the efficiency of the proposed solution against the \textit{UWF-2022} dataset containing records from 2022 specifically created to aid detection and classification of network attacks.

\subsection{Dataset~UWF-2022}\label{subsec:dataset}
The \textit{UWF-2022} dataset comprises several files with the extension \textit{.parquet} \cite{parquet-docs} that correspond to enhanced versions of traditional \textit{.csv} files.
Given the complex nature of this dataset, a two-pronged classification approach was necessary: (1) collating all source files together into a cohesive dataset and (2) conducting malware classification.
After parsing and filtering source files to eliminate non-essential inputs for this research, a structured dataset was produced, with its attributes listed in Table \ref{tab:uwfattr}.

\begin{table}[ht]
	\begin{threeparttable}
		\caption{Dataset UWF-2022 Attributes}
		\label{tab:uwfattr}
		\setlength\tabcolsep{0pt} 
		
		\begin{tabular*}{\columnwidth}{@{\extracolsep{\fill}} lll}
			\toprule
			Attribute Name\tnote{a} & Type\tnote{b} & Description\tnote{c} \\
			\midrule
			resp\textunderscore pkts & int32 & Number of IP level bytes responder sent\\
			service & object & Application layer protocol\\
			orig\textunderscore ip\textunderscore bytes & int32 & Number of packet responder sent \\
			local\textunderscore resp &  bool & \makecell[l]{Whether the connection is\\responded locally}\\
			missed\textunderscore bytes & int32 & Representation of packet loss\\
			proto & object & Transport layer protocol of connection\\
			duration & float64 & How long connection lasted\\
			conn\textunderscore state & object & Possible connection state\\
			dest\textunderscore ip\textunderscore zeek & object & Destination IP address\\
			orig\textunderscore pkts & int32 & Number of packet originator sent\\
			resp\textunderscore ip\textunderscore bytes & int32 & Number of IP level bytes responder sent\\
			dest\textunderscore port\textunderscore zeek & int32 & Destination Port\\
			orig\textunderscore bytes & float64 & Number of payload bytes originator sent\\
			local\textunderscore orig & boolean & Whether the originator is a local user\\
			datetime & datetime64 & \makecell[l]{Timestamp when the packet was\\processed}\\
			resp\textunderscore bytes & float64 & Number of bytes responder sent\\
			src\textunderscore port\textunderscore zeek & int32 & Source port\\
			ts & float64 & Time of first packet\\
			src\textunderscore ip\textunderscore zeek & object & Source IP address\\
			
			\addlinespace
			label\textunderscore tactic & int64 & Output class. It classifies the malware\\
			
			\bottomrule
		\end{tabular*}
		
		\smallskip
		\scriptsize
		\begin{tablenotes}
			\RaggedRight
			\item[a] Name of the input class.
			\item[b] Original Data type of the input.
			\item[c] Original description from the dataset creators.
		\end{tablenotes}
	\end{threeparttable}
\end{table}

Table \ref{tab:uwfattr} clearly displays all of the attributes the machine learning algorithm will examine: they include common network parameters and metrics data that enable comparison with others in future analyses. An advantage of using machine learning for threat hunting lies in its model adaptability - real-time training can take place using data collected via network sniffers followed by batch learning algorithms allowing real-time training of models afterwards.

Effective classification of malware is key in this project, although no universally agreed-upon classification method exists; most research centers utilize MITRE's ATT\&CK matrix as their standard classification system for possible output classes; see Table \ref{tab:uwfclasees} for this list of possible output classes. Utilizing such an effective classification system during machine learning process helps in efficiently categorizing malware while serving as cross reference among datasets or research projects.

\begin{table}[ht]
	\begin{threeparttable}
		\caption{Dataset UWF-2022 Classes}
		\label{tab:uwfclasees}
		\setlength\tabcolsep{0pt} 
				
		\begin{tabular*}{\columnwidth}{@{\extracolsep{\fill}} lll}
			\toprule
			Class Name\tnote{a} & \# of Samples\tnote{b} & MITRE TACTIC\tnote{c} \\
			\midrule
			Benign traffic & 9,281,599 & Desired traffic\\
			Reconnaissance & 9,278,722 & \makecell[l]{T1590, T1592, T1595,\\T1595.001, and T1595.002}\\
			Discovery & 2,086 & T1046, and T1135\\
			Credential Access & 31 & \makecell[l]{T1003.002, T1003.008,\\T1110, and T1552}\\
			Privilege Escalation & 13 & T1068\\
			Exfiltration & 7 & T1048.001\\
			Lateral Movement & 4 & T1021, and T1021.004\\
			Resource Development & 3 & T1587.004, and T1588.002\\
			Defense Evasion & 1 & \makecell[l]{T1070, T1070.002,\\T1070.003, and T1564}\\
			Initial Access & 1 & T1189, and T1190\\
			Persistence & 1 & T1098, T1136, and T1136.001\\
		\end{tabular*}
				
		\smallskip
		\scriptsize
		\begin{tablenotes}
			\RaggedRight
			\item[a] Name of the class in the UWF-2022 dataset.
			\item[b] Number of samples for that class.
			\item[c] The label of the tactic in the MITRE ATT\&CK classification.
		\end{tablenotes}
	\end{threeparttable}
\end{table}

Table \ref{tab:uwfclasees} makes clear this dataset is quite unbalanced in its sample of attack classes; thus making the learning process significantly harder. Still, this reflects real life as certain attacks tend to be more likely.

Besides, Table \ref{tab:uwfclasees} clearly illustrates that this dataset exhibits an imbalance of classes; certain have greater representation than others, potentially creating obstacles during training as machine learning tends to favor majority classes more and therefore provide less accurate predictions for minority ones.

Additionally, it should be acknowledged that this dataset accurately represents real networks. As seen in reality, some types of attacks tend to be more prevalent and probable than others - which mirrors an inherent imbalance in threats and intrusions within networks - making the dataset an accurate representation of actual network behavior.

However, another classification strategy dividing traffic into only two classes - beneficial and malicious traffic - may produce some measure of equilibrium while making learning significantly more complex.
Complexity arises due to being unable to predict or gauge the severity or magnitude of potential attacks; creating uncertainty for learning endeavors.

So as to take into account this complex aspect, we planned a deliberate course of action, in which experimental pursuits would be organized on an inherently imbalanced dataset.

\subsubsection{Dealing with an unbalance dataset}\label{subsec:uwfunbalanced}

The challenge of Threat hunting involves uncovering less likely attacks, as these tend to be among the most complex and harmful. Such attacks could even serve as precursors of APTs. Given an unevenly distributed dataset, class imbalance must also be addressed if one wishes to create an accurate predictive model that works reliably over time.

To address this issue, the initial step should be removing classes with only one instance. While this may not be ideal, removing these classes helps avoid biased learning and may improve predictive model performance by prioritizing more significant classes, moreover, by removing classes with minimal representation, it is possible to ensure a more balanced distribution.

Step two requires developing an algorithm to reduce instances in over-represented classes and achieve more equitable distribution among other classes, to counteract any skew caused by over-represented instances in certain categories. It is critical that this step be completed as it helps counteract any imbalance caused by an disproportionate distribution.
The algorithm for balancing the dataset can be outlined in three steps:

\begin{itemize}
    \item \textbf{Undersampling}: By eliminating instances from the majority class, your dataset becomes more representative of all classes, which allows your model to learn from and predict more accurately from all.
    \item \textbf{Random oversampling}: Oversampling refers to artificially increasing instances in minority classes by simulating random existing data sets and replicating it; in doing this, oversampling will give more representation for underrepresented classes within a dataset, providing better exposure for models targeting these specific classes..
    \item \textbf{SMOTE oversampling}: Similar to random oversampling, synthetic oversampling generates data points by interpolating among existing instances of minority classes instead of duplicating them directly.
\end{itemize}

By following these steps, the dataset becomes more balanced, enabling its model to focus on both prevalent and less likely attacks, increasing its ability to effectively recognize and classify threats.
Table \ref{tab:uwfbalance} allows viewers to track the growth of data set while applying balancing algorithm.

To maximize accuracy in machine learning algorithms, addressing dataset imbalance is vitally important. Unfortunately, due to vast disparities between majority and minority classes in this dataset, equal representation across classes would not be practicable or realistic. Oversampling may provide some benefits while simultaneously increasing risks.

Oversampling may help overcome difficulties caused by imbalanced data; however, overfitting can impede generalization capacity of models and require further consideration to find an ideal approach to handling this class imbalance based on your dataset's specific characteristics and challenges.

\begin{table}[ht]
	\begin{threeparttable}
		\caption{Evolution of the Dataset UWF-2022 Samples During the Balancing Process}
		\label{tab:uwfbalance}
		\setlength\tabcolsep{0pt} 
						
		\begin{tabular*}{\columnwidth}{@{\extracolsep{\fill}} lllll}
			\toprule
			Class Name\tnote{a} & Initial\tnote{b} & Under\tnote{c} & Over\tnote{d} & SMOTE\tnote{e} \\
			\midrule
			Benign traffic & 9,281,599 & 8,874 & 8,874 & 8,874\\
			Reconnaissance & 9,278,722 & 9,176 & 9,176 & 9,176\\
			Discovery & 2,086 & 2,086 & 2,086 & 2,086\\
			\makecell[l]{Credential\\Access} & 31 & 31 & 70 & 70 \\
			\makecell[l]{Privilege\\Escalation} & 13 & 13 & 130 & 226\\
			Exfiltration & 7 & 7 & 70 & 166 \\
			\makecell[l]{Lateral Movement} & 4 & 4 & 40 & 136\\
			\makecell[l]{Resource\\Development} & 3 & 3 & 30 & 126\\
			\makecell[l]{Defense\\Evasion} & \multicolumn{4}{c}{Removed} \\
			\makecell[l]{Initial\\Access} & \multicolumn{4}{c}{Removed} \\
			Persistence & \multicolumn{4}{c}{Removed} \\
		\end{tabular*}
						
		\smallskip
		\scriptsize
		\begin{tablenotes}
			\RaggedRight
			\item[a] Name of the class in the UWF-2022 dataset.
			\item[b] Initial number of samples for that class.
			\item[c] Number of samples after applying undersampling.
			\item[d] Number of samples after applying random oversampling.
			\item[e] Number of samples after applying SMOTE oversampling. This is the final number of samples as well.
		\end{tablenotes}
	\end{threeparttable}
\end{table}

\subsubsection{Training the model}\label{subsec:uwftraining}

After successfully making our dataset more conducive for machine learning algorithms by correcting class imbalance, conducting extensive research on suitable classification algorithms for log data, and selecting \textit{random forest} classifier as our decision, extensive investigations indicate it provides superior accuracy when handling such classification challenges \cite{Sharafaldin2018TowardGA}, \cite{app122110761}, and \cite{9671731}.

In order to undertake an adequate evaluation, an elegant machine learning algorithm was meticulously created using Python libraries. Once all preparatory stages have been successfully executed, Threat Trekker will accept one parquet file as its input before performing an appropriate transformation to produce a pandas dataframe and enable subsequent operations.

Engaging the procedural intricacies, the algorithm carefully splits up the dataframe into its constituent parts and assigns 70\% for training; 30\% is kept back as reserve data to serve as testing material later.

An essential aspect is how input attributes are encoded prior to being introduced into the training process. Particularly notable, categorical fields like connection state and protocol type were encoded using One Hot Encoding; similarly recurring attributes used as algorithm inputs include IPv4 addresses which may be encoded using various techniques detailed by this thesis \cite{Shao2019}.

Given the size and scope of this classification task, data must be encoded efficiently to be meaningfully classified. A parsing algorithm was adopted for IP address strings to efficiently convert them to singular integers through segmenting the IP address into four eight-bit numbers and concatenating them together for one 32-bit value; Algorithm \ref{alg:ipv4_to_int32} clearly depicts this.

\begin{algorithm}[H]
	\caption{Encoding an IPv4 Address Into a 32-bit Integer}
	\label{alg:ipv4_to_int32} 
	\begin{algorithmic}[1]
		\Procedure{ipv4\textunderscore to\textunderscore int32}{ipv4}
		\State split $\gets$ \Call{StringSplit}{ipv4, "."}
		\State result $\gets$ (\Call{ToInt}{split[0]} $\ll$ 24) 
		\\ \hspace{3em}  + (\Call{ToInt}{split[1]} $\ll$ 16)
		\\ \hspace{3em}  + (\Call{ToInt}{split[2]} $\ll$ 8)
		\\ \hspace{3em}  + \Call{ToInt}{split[3]}
		\State \Return result
		\EndProcedure
	\end{algorithmic}
\end{algorithm}

Also, all timestamps are converted to Unix Epoch format to maintain consistency across attributes and ensure coherency across data types. Furthermore, since all remaining attributes are numeric values, automatic casting to float becomes possible.

\subsubsection{Consolidation of all attack variants for binary classification}\label{subsec:uwftwofold}

An alternative approach for dataset balancing involves grouping all attack variants together into a unified category and configuring an algorithm for binary classification, creating two potential output classes -- beneficial or malicious log samples. Following preprocessing using this approach, both output classes achieved near equal probabilities as evidenced in Table \ref{tab:uwfclaseestwofold}, creating optimal balance within the dataset without additional adjustments being necessary.

\begin{table}[ht]
	\begin{threeparttable}
		\caption{Dataset UWF-2022 Classes in a Twofold Representation}
		\label{tab:uwfclaseestwofold}
		\setlength\tabcolsep{0pt} 
				
		\begin{tabular*}{\columnwidth}{@{\extracolsep{\fill}} lll}
			\toprule
			Class Name\tnote{a} & \# of Samples\tnote{b} & MITRE TACTIC\tnote{c} \\
			\midrule
			Benign traffic & 9.178.959 & Desired traffic\\
			Malicious & 8.874.034 & Amalgamated\\
		\end{tabular*}
				
		\smallskip
		\scriptsize
		\begin{tablenotes}
			\RaggedRight
			\item[a] Name of the class in the UWF-2022 dataset.
			\item[b] Number of samples for that class.
			\item[c] The label of the tactic in the MITRE ATT\&CK classification.
		\end{tablenotes}
	\end{threeparttable}
\end{table}

While the preceding approach is anticipated to yield exceptionally high precision, it is imperative to acknowledge that it entails a level of generalization across the various attacks. Consequently, the ability to discern the least probable attacks becomes constrained within this framework.

While this approach promises exceptional precision, it must also be acknowledged that its generalized approach may prevent detection of less likely attacks within its framework; hence its ability to identify persistence hazards is reduced significantly within this framework.

\subsection{Dataset~CIC-IDS2017}\label{subsec:dataset-cicids}
CIC-IDS2017 dataset was specifically tailored to detect network attacks. As it encompasses multiple network capture files that were parsed into CSV files, this dataset features significantly higher numbers of input classes compared with UWF-2022 dataset. To maintain consistency between all datasets regarding input attributes across them all, some original columns had to be cropped off of CIC-IDS2017. The attributes used to train the model are described in table \ref{tab:cicattr}.

\begin{table}[ht]
	\begin{threeparttable}
		\caption{Dataset CIC-IDS2017 Attributes}
		\label{tab:cicattr}
		\setlength\tabcolsep{0pt} 
		\begin{tabular*}{\columnwidth}{@{\extracolsep{\fill}} l@{\hspace{0.3em}} l@{\hspace{0.3em}} l} 
			\toprule
			Attribute Name\tnote{a} & Type\tnote{b} & Description\tnote{c} \\
			\midrule
			Destination Port & int64 & Port number on the destination system\\
			Flow Duration & int64 & Duration of the communication in ms\\
			Total Fwd Packets & int64 & Packets sent in the forward direction\\
			Total Backward Packets &  int64 & Packets sent in the backward direction\\
			\makecell[l]{Total Length Of\\Fwd Packets} & int64 & \makecell[l]{Total size of packets sent in the\\forward direction during the flow}\\
			\makecell[l]{Total Length Of\\Bwd Packets} & int64 & \makecell[l]{Total length size of packets sent in the\\backward direction during the flow} \\
			Flow Bytes/s  & float64 & Bytes sent per sec. in the flow (avg.) \\
			Flow Packets/s  & float64 & Packets sent per sec. in the flow (avg.)\\
			Fwd Header Length  & int64 & Length of the headers sent forward \\
			Bwd Header Length  & int64 & Length of the received headers\\
			Fwd Packets/s & float64 & \makecell[l]{The avg. number of packets sent\\per second in the forward direction}\\
			Bwd Packets/s & float64 & \makecell[l]{The avg. number of packets sent\\per second in the backward direction}\\
			Min Packet Length & int64 & The min. length of a packet in the flow\\
			Max Packet Length & int64 & The max. length of a packet in the flow\\
			Avg Fwd Segment Size & float64 & The avg. number of sent segments \\
			Avg Bwd Segment Size & float64 & The avg. number of received segments \\
			\makecell[l]{Init Win Bytes\\Forward} & int64 & \makecell[l]{Initial window size in the\\forward direction}\\
			\makecell[l]{Init Win Bytes\\Backward} & int64 & \makecell[l]{Initial window size in the\\backward direction}\\
			Act Data Pkt Fwd & int64 & \makecell[l]{The total number of actual data\\packets sent in the forward direction}\\
			Min Seg Size Forward & int64 & \makecell[l]{The minimum segment size observed\\in the forward direction}\\
			   
			\addlinespace
			Label & object & Output class. It classifies the malware\\
						
			\bottomrule
		\end{tabular*}
				
		\smallskip
		\scriptsize
		\begin{tablenotes}
			\RaggedRight
			\item[a] Name of the input attributes.
			\item[b] Original data type of the attribute.
			\item[c] Description of the attributes.
		\end{tablenotes}
	\end{threeparttable}
\end{table}

At first glance, CIC-IDS2017 features attributes primarily consist in network metrics that may easily be extracted from any form of communication between hosts. This decision stems from Threat Trekker's insistence upon generality allowing it to work efficiently across diverse network environments regardless of their characteristics or nature.

Furthermore, it should be noted that CIC-IDS2017 dataset contains fourteen distinct output classes with more balanced distribution compared to UWF-2022 dataset (detailed below in table \ref{tab:cicclasees}). Therefore, no class had to be removed as all fourteen classes remained minimally represented and retained for analysis.

\begin{table}[ht]
	\begin{threeparttable}
		\caption{Dataset CIC-IDS2017 Classes}
		\label{tab:cicclasees}
		\setlength\tabcolsep{0pt} 
						
		\begin{tabular*}{\columnwidth}{@{\extracolsep{\fill}} l@{\hspace{0.3em}} l@{\hspace{0.3em}} l}
			\toprule
			Class Name\tnote{a} & \# of Samples\tnote{b} & MITRE TACTIC\tnote{c} \\
			\midrule
			Benign                   & 2,273,097 & Desired traffic\\
			DoS Hulk                 & 231,073 & T1498\\
			PortScan                 & 158,930 & T1046\\
			DDoS                     & 128,027 & T1498, T1498.002\\
			DoS GoldenEye            & 10,293 & T1498\\
			FTP-Patator              & 7,938 & T1110, T1114\\
			SSH-Patator              & 5,897 & T1110, T1114\\
			DoS slowloris            & 5,796 & T1498\\
			DoS Slowhttptest         & 5,499 & T1498\\
			Bot                      & 1,966 & T1508\\
			Web Attack Brute Force   & 1,507 & T1110\\
			Web Attack XSS           & 652 & T1059\\
			Infiltration             & 36 & T1566\\
			Web Attack Sql Injection & 21 & T1110\\
			Heartbleed               & 11 & T1504\\
		\end{tabular*}
						
		\smallskip
		\scriptsize
		\begin{tablenotes}
			\RaggedRight
			\item[a] Name of the class in the CIC-IDS2017 dataset.
			\item[b] Number of samples for that class.
            \item[c] The label of the tactic in the MITRE ATT\&CK classification.
		\end{tablenotes}
	\end{threeparttable}
\end{table}

Though algorithms often perform commendably, dataset balancing remains an excellent practice to enhance comparative analyses and facilitate greater insight. The main goal of dataset balancing is to improve its suitability for machine learning applications by decreasing instances within majority classes; with particular attention focused on mitigating any overfitting phenomena that might otherwise arise from excessive classification accuracy.

Table \ref{tab:cicbalance} offers an exhaustive overview of all changes effected by each iteration of the balancing algorithm to the dataset, serving as an essential reference document and showing stepwise transformation executed to reach equilibrium among class instances.

\begin{table}[ht]
	\begin{threeparttable}
		\caption{Evolution of the Dataset CIC-IDS2017 Samples During the Balancing Process}
		\label{tab:cicbalance}
		\setlength\tabcolsep{0pt} 
								
		\begin{tabular*}{\columnwidth}{@{\extracolsep{\fill}} lllll}
			\toprule
			Class Name\tnote{a} & Initial\tnote{b} & Under\tnote{c} & Over\tnote{d} & SMOTE\tnote{e} \\
			\midrule
			Benign                   & 2,273,097 & 45,426 & 45,426 & 45,426\\
			DoS Hulk                 & 231,073 & 23,012 & 23,012 & 23,012\\
			PortScan                 & 158,930 & 15,880 & 15,880 & 15,880\\
			DDoS                     & 128,027 & 12,802 & 12,802 & 12,802\\
			DoS GoldenEye            & 10,293 & 10,293 & 10,293 & 10,293\\
			FTP-Patator              & 7,938 & 7,935 & 7,935 & 8,149\\
			SSH-Patator              & 5,897 & 5,897 & 5,897 & 6,111\\
			DoS slowloris            & 5,796 & 5,796 & 5,796 & 6,010\\
			DoS Slowhttptest         & 5,499 & 5,499 & 5,499 & 5,713\\
			Bot                      & 1,966 & 1,956 & 1,956 & 2,170\\
			\makecell[l]{Web Attack\\Brute Force}   & 1,507 & 1,507 & 1,507 & 1,721\\
			Web Attack XSS           & 652 & 652 & 1,304 & 1,518\\
			Infiltration             & 36 & 36 & 360 & 574\\
			\makecell[l]{Web Attack\\Sql Injection} & 21 & 21 & 210 & 424\\
			Heartbleed               & 11 & 11 & 110 & 324\\
		\end{tabular*}
								
		\smallskip
		\scriptsize
		\begin{tablenotes}
			\RaggedRight
			\item[a] Name of the class in the CIC-IDS2017 dataset.
			\item[b] Initial number of samples for that class.
			\item[c] Number of samples after applying undersampling.
			\item[d] Number of samples after applying random oversampling.
			\item[e] Number of samples after applying SMOTE oversampling. This is the final number of samples as well.
		\end{tablenotes}
	\end{threeparttable}
\end{table}


\subsection{Dataset~TON-IoT-2019}\label{subsec:dataset-toniot}

As it is shown by our analysis of previous datasets, machine learning efforts are dramatically affected by data distribution skew. Such irregularity creates challenges to reaching results that meet robust reliability thresholds; hence this section investigates performance characteristics of Threat Trekker Algorithm when applied to the TON\textunderscore IoT dataset which features only two output classes; benign traffic and malicious traffic.

TON\textunderscore IoT dataset comprises logs derived from both contemporary IoT devices and industrial IIoT (IIoT) systems, used as network metrics in model training. Table \ref{tab:tonattr} details these attributes thoroughly.

\begin{table}[ht]
	\begin{threeparttable}
		\caption{Dataset TON\textunderscore IoT Attributes}
		\label{tab:tonattr}
		\setlength\tabcolsep{0pt} 
		\begin{tabular*}{\columnwidth}{@{\extracolsep{\fill}} l@{\hspace{0.3em}} l@{\hspace{0.3em}} l} 
			\toprule
			Attribute Name\tnote{a} & Type\tnote{b} & Description\tnote{c} \\
			\midrule
			src\textunderscore ip & object & Source IP addresses\\
			src\textunderscore port & int64 & Source ports (TCP/UDP) \\
			dst\textunderscore ip & object & Destination IP addresses\\
			dst\textunderscore port &  int64 & Destination ports (TCP/UDP)\\
			proto & object & Transport layer protocols of flow connections\\
			duration & float64 &  The time of the packet connections\\
			src\textunderscore bytes  & int64 & \makecell[l]{Source bytes which are originated from\\payload bytes of TCP sequence numbers}\\
			dst\textunderscore bytes  & int64 & \makecell[l]{Destination bytes which are responded\\payload bytes from TCP sequence numbers}\\
			conn\textunderscore state  & object &  Various connection states\\
			src\textunderscore pkts   & int64 & \makecell[l]{Number of original packets which is\\estimated from source systems}\\
			src\textunderscore ip\textunderscore bytes & int64 & \makecell[l]{Total length of IP header field of\\source systems}\\
			dst\textunderscore pkts & int64 & \makecell[l]{Number of destination packets which is\\estimated from destination systems}\\
			dst\textunderscore ip\textunderscore bytes  & int64 & \makecell[l]{Total length of IP header field of\\destination systems}\\
						
			\addlinespace
			label & object & Whether traffic is benign or malicious\\
									
			\bottomrule
		\end{tabular*}
						
		\smallskip
		\scriptsize
		\begin{tablenotes}
			\RaggedRight
			\item[a] Name of the input attributes.
			\item[b] Original data type of the attribute.
			\item[c] Description of the attributes.
		\end{tablenotes}
	\end{threeparttable}
\end{table}

Reiterating my earlier point, the TON\textunderscore IoT dataset clearly delineated two output classes within its data set - benign traffic being twofold more common compared to instances of malicious traffic. Table \ref{tab:tonclaseesunbalanced} offers further clarification for this statement.

\begin{table}[ht]
	\begin{threeparttable}
		\caption{Dataset TON\textunderscore IoT Classes}
		\label{tab:tonclaseesunbalanced}
		\setlength\tabcolsep{0pt} 
								
		\begin{tabular*}{\columnwidth}{@{\extracolsep{\fill}} l@{\hspace{0.3em}} l@{\hspace{0.3em}} l}
			\toprule
			Class Name\tnote{a} & \# of Samples\tnote{b} & MITRE TACTIC\tnote{c} \\
			\midrule
			Benign                   & 300,000 & Desired traffic\\
			Malicious                & 161,043 & Amalgamated\\
		\end{tabular*}
								
		\smallskip
		\scriptsize
		\begin{tablenotes}
			\RaggedRight
			\item[a] Name of the class in the CIC-IDS2017 dataset.
			\item[b] Number of samples for that class.
			\item[c] The label of the tactic in the MITRE ATT\&CK classification.
		\end{tablenotes}
	\end{threeparttable}
\end{table}

As there was already a significant reservoir of original malicious samples present in the authentic dataset, and as its authors had undersampled majority classes already, an informed decision was made not to balance this dataset further.

As we now appreciate that precision in classification can approach perfection, our next step involves undertaking an intricate task: identifying specific classes of attack similar to our prior datasets. Instead of binary division, this step prepares us for nuanced differentiation of output classes - Table \ref{tab:tonclasses} offers more granular insight into this complex categorization process.

\begin{table}[ht]
	\begin{threeparttable}
		\caption{Dataset TON\textunderscore IoT Classes}
		\label{tab:tonclasses}
		\setlength\tabcolsep{0pt} 
								
		\begin{tabular*}{\columnwidth}{@{\extracolsep{\fill}} l@{\hspace{0.3em}} l@{\hspace{0.3em}} l l}
			\toprule
			Class Name\tnote{a} & \makecell[l]{Original \# of\\ Samples (PCAP)}\tnote{b} & \makecell[l]{\# of Samples\\(experiments)}\tnote{c} & MITRE TACTIC\tnote{d} \\
			\midrule
			Benign            & 796,380    & 300,000 & Desired traffic\\
			Scanning          & 7,140,161  & 20,000  & T1595\\
			DoS               & 3,375,328  & 20,000  & T1498\\
			Injection         & 452,659    & 20,000  & T1203 \\
			Ddos              & 6,165,008  & 20,000  & T1498, T1498.002\\
			Password          & 1,718,568  & 20,000  & T1110\\
			Xss               & 2,108,944  & 20,000  & T1189\\
			Ransomware        & 72,805     & 20,000  & T1486\\
			Backdoor          & 508,116    & 20,000  & T1098\\
			Mitm              & 1,052      & 1,043   & T1557\\
		\end{tabular*}
								
		\smallskip
		\scriptsize
		\begin{tablenotes}
			\RaggedRight
			\item[a] Name of the class in the TON\textunderscore IoT dataset.
            \item[b] Number of samples originally recovered for that class in the traffic captures.
			\item[c] Undersampled number of samples for that class (Actual number of samples which are used for performing this experiment).
			\item[d] The label of the tactic in the MITRE ATT\&CK classification.
		\end{tablenotes}
	\end{threeparttable}
\end{table}

It is evident that samples across most output classes have an even distribution, due to the data set's authors' careful use of an undersampling approach applied uniformly across all classes. This strategy increases minority exposure once balanced out; providing them with more opportunities and increasing likelihood for exposure once balanced out.

Noteworthy is the success of this singular dataset's balancing process in particular. This can be explained by its inherent effectiveness due to the significance of minority classes within it; accordingly, refraining from oversampling seems appropriate given their significant presence within this sample.
Balance within the dataset has resulted in an impressive result: an assertion regarding its greater precision compared to others is justifiable.

%% file: sections/results.tex
The efforts undertaken in the antecedent sections have revealed the feasibility of parsing, aggregating, and filtering log data in order to make it more suitable for machine learning algorithms.

Therefore, this discourse aims at providing an in-depth comparison of results achieved via implementing a machine learning regimen utilizing Random Forest classifier.

\subsection{Dataset~UWF-2022}\label{subsec:resuwf}

\begin{figure}[ht]
  \centering
  \includegraphics[width=\columnwidth]{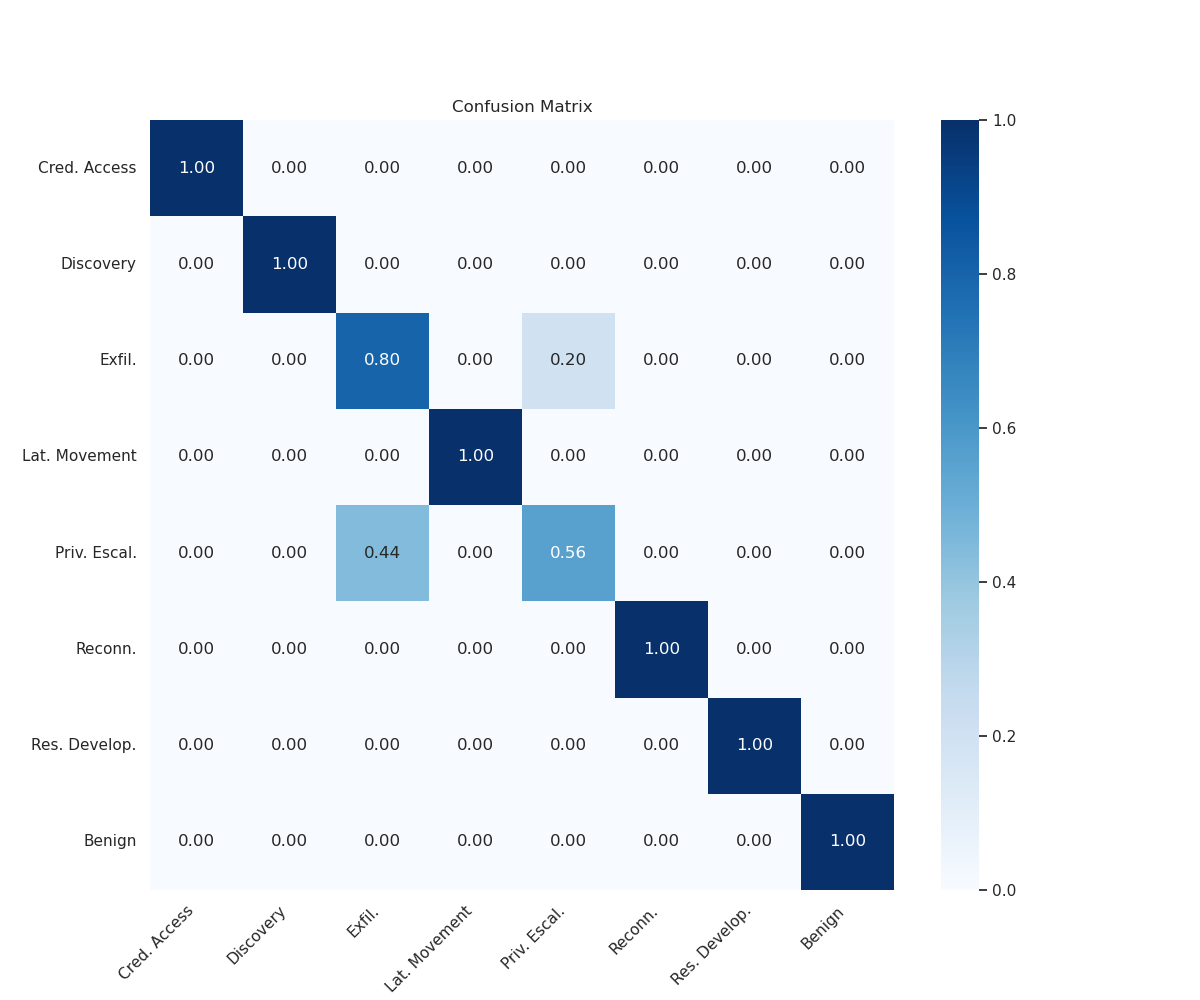}
  \caption{Dataset UWF-2022: Confusion Matrix Balance}
  \label{fig:uwfcm}
\end{figure}

Figure \ref{fig:uwfcm} depicts the outcomes of the model training endeavor succinctly. From its depiction, it is evident that its algorithm boasts a notable degree of accuracy when making predictions; however, there are two classes where its predictive capabilities fall short of expectations.

This discrepancy demonstrates the difficulties associated with class imbalance and correctly classifying certain threats. Even though achieving high accuracy may be achieved overall, model performance may suffer when dealing with less represented or more complex classes.

Threat Trekker was examined as part of its ability to effectively manage data samples with inherent asymmetry, effectively handling any imbalance. Figure \ref{fig:uwfcm} provides evidence of notable achievements achieved via Threat Trekker; such achievements included high macro average scores. Nonetheless, for comprehensive evaluation and comparison purposes these outcomes shall also be contrasted against results obtained using others datasets.

\subsubsection{Utilizing binary classification for output class categorization}\label{subsec:binaryclassification}

\begin{figure}[ht]
  \centering
  \includegraphics[width=\columnwidth]{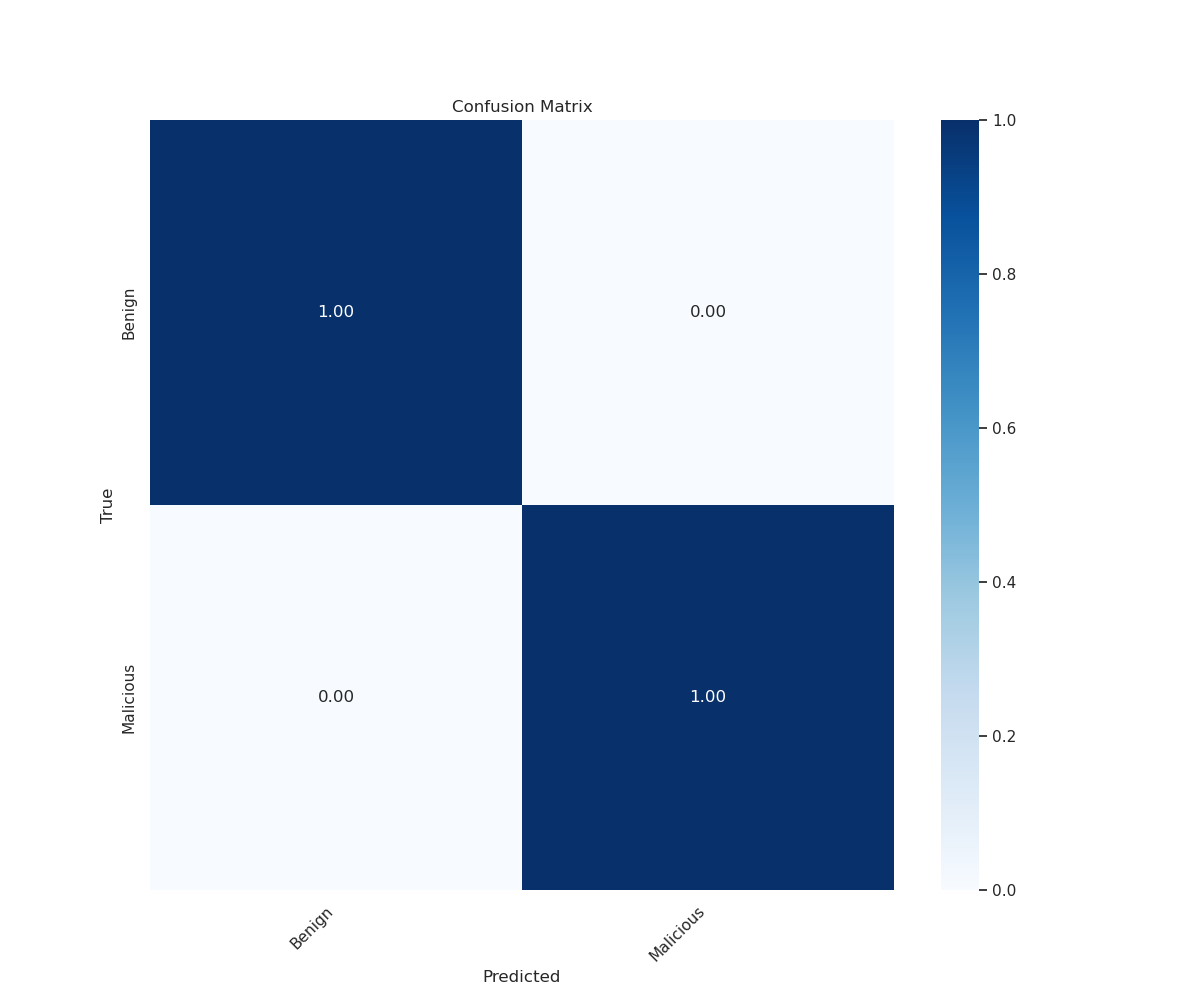}
  \caption{Dataset UWF-2022: Confusion Matrix in a Twofold Representation}
  \label{fig:uwfcmtwofold}
\end{figure}

Figure \ref{fig:uwfcmtwofold} displays impressive precision when amalgamating attack classes; this experiment shows the viability of employing machine learning algorithms for log classification given an extensive and authentic dataset, thus mitigating overfitting risks.

\subsection{Dataset~CID-IDS2017}\label{subsec:rescic}

\begin{figure}[ht]
  \centering
  \includegraphics[width=\columnwidth]{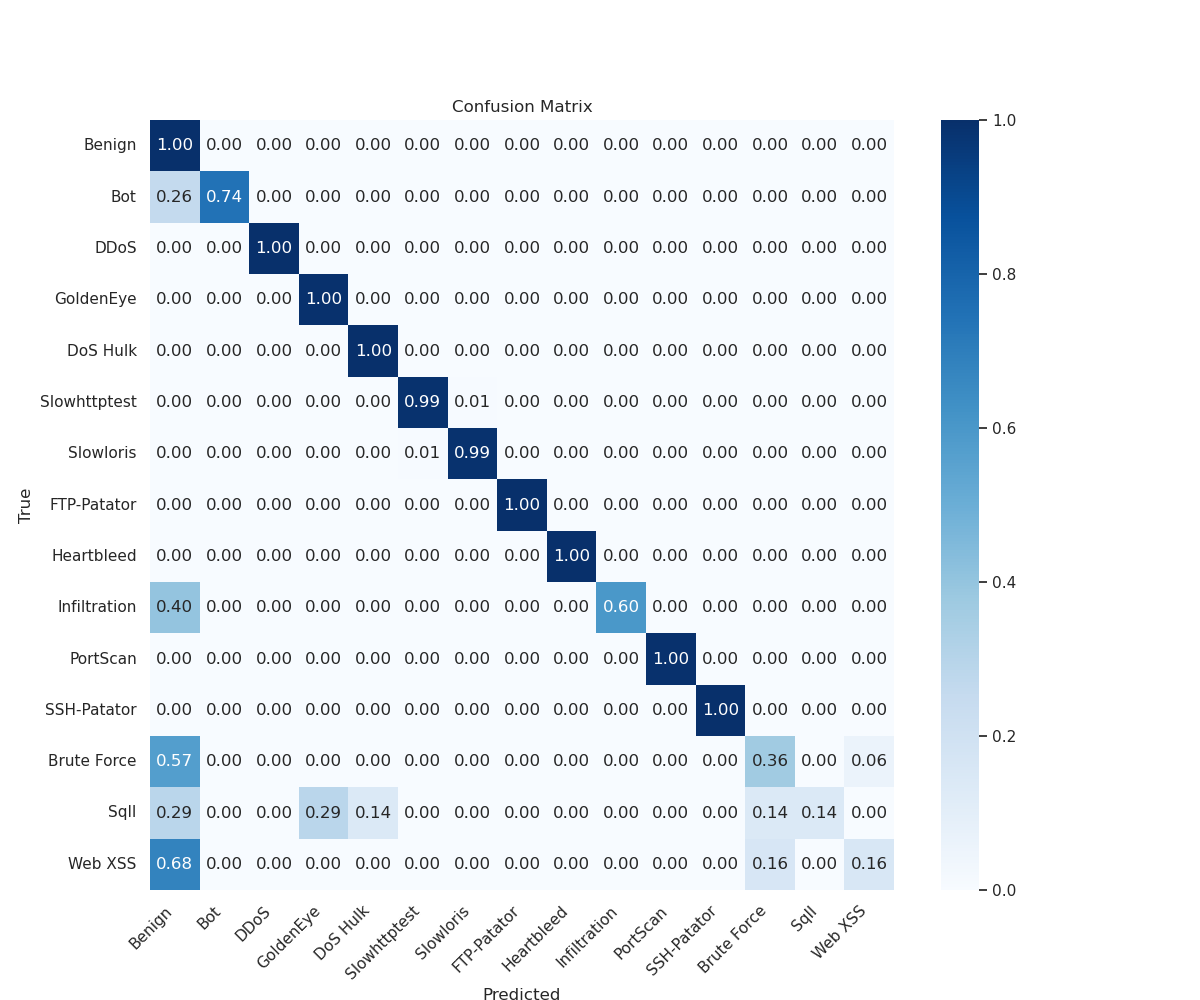}
  \caption{CIC-IDS2017: Confusion Matrix}
  \label{fig:cicunmbalance}
\end{figure}

After training the algorithm using the CIC-IDS2017 dataset without explicit balancing of said dataset, an unexpected but remarkable outcome emerged: its precision was impressively high despite comprising fourteen separate classes in its composition. Figure  \ref{fig:cicunmbalance} vividly displays this outcome: in most output classes and two outlier instances it achieved pinpoint accuracy for output classes.

Although the algorithm's results may seem encouraging, they must also take into account that their success stemmed from being trained on an inherently disparate dataset.

\begin{figure}[ht]
  \centering
  \includegraphics[width=\columnwidth]{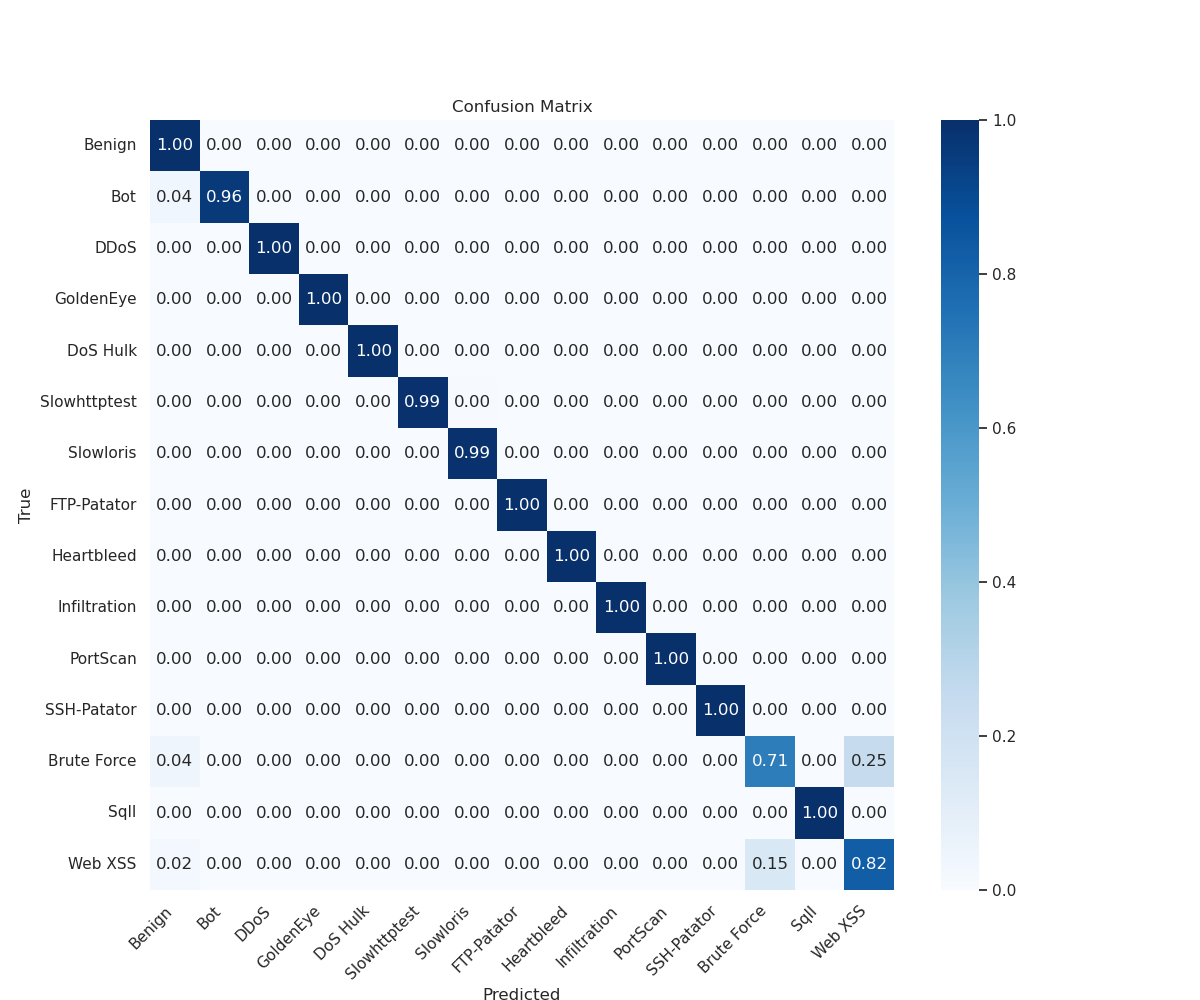}
  \caption{CIC-IDS2017: Balanced Confusion Matrix}
  \label{fig:cicmbalance}
\end{figure}

Figure \ref{fig:cicmbalance} provides an eye-opening snapshot of the results following successful dataset rebalancing. Evident in its depiction is how closely similar its performance outcomes pre and post-balancing were. Notably, an obvious distinction emerges in class situations in which algorithm performance shows reduced efficacy. This discrepancy can be explained by a significant reduction in sample representation from major classes as a direct result of balancing procedures. By doing this, cases from less frequent vulnerabilities gain increased influence upon probability estimation overall.

\subsection{Dataset~TON-IoT-2019}\label{subsec:reston}

An extraordinary finding was revealed upon training the model with the TON\textunderscore IoT dataset: converging all attacks types into one category led to an immediate boost in model accuracy.
Strategic consolidation led to more equitable distribution of classes within malicious traffic groups, improving balance vis-a-vis their benign counterpart.

\begin{figure}[ht]
  \centering
  \includegraphics[width=\columnwidth]{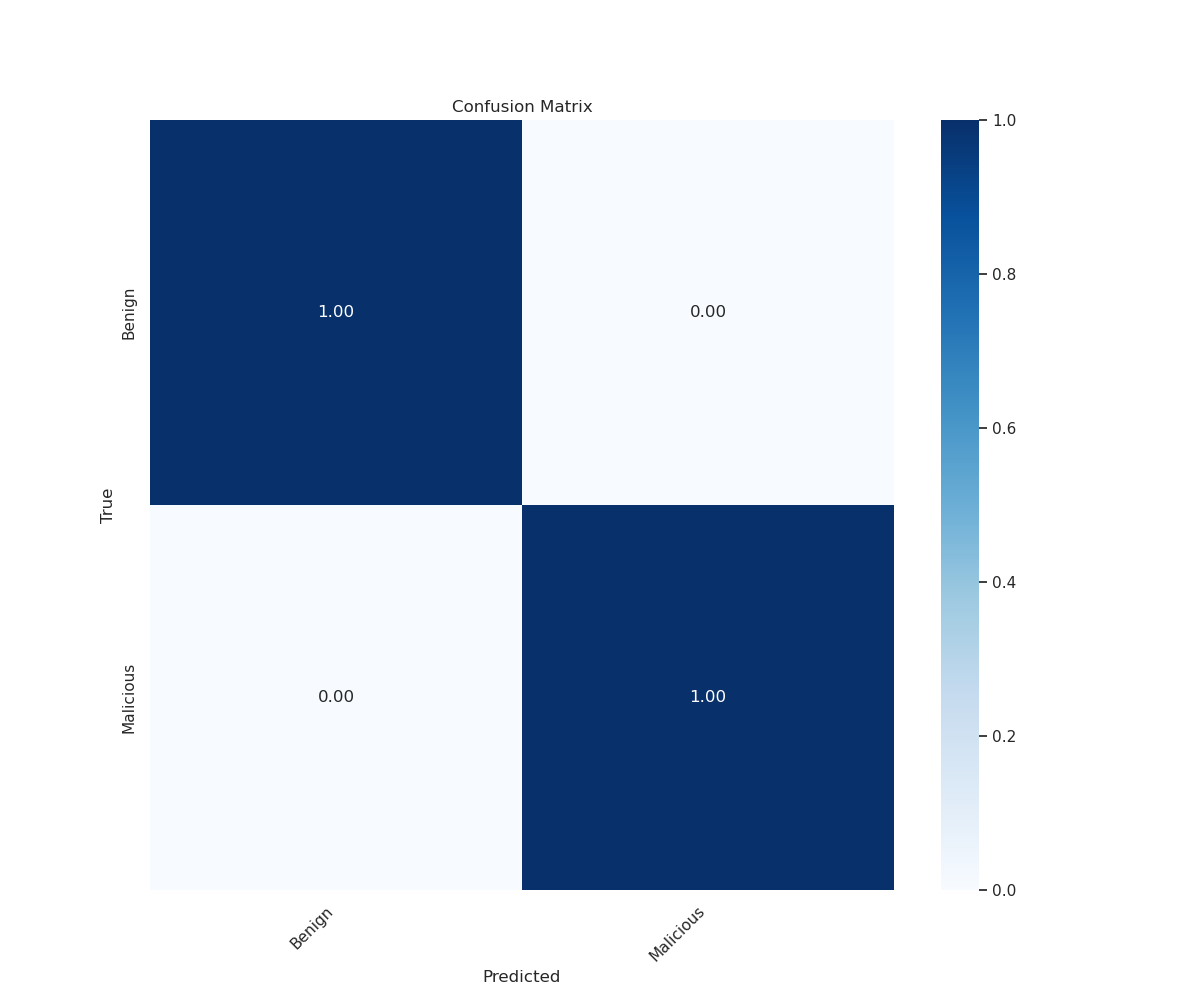}
  \caption{TON\textunderscore IoT: Confusion Matrix - Twofold Segregation}
  \label{fig:tonunmbalance}
\end{figure}

Figure \ref{fig:tonunmbalance} depicts graphically how the model's accuracy trajectory achieved excellence and reached perfection, noting also how this result could be replicated through implementation of output class balancing procedures.

This approach has effectively demonstrated the ability of Random Forest classifiers in accurately classifying log data with minimal error, yielding results with high precision. Unfortunately, such forms of analysis often veer away from their initial intention - which was to perform fine-grained classification of network attacks so as to detect and defend against those incidents which represented the greatest risks to network integrity.

\subsubsection{Employing an explicit label for each category of attack}\label{subsec:grainedton}

\begin{figure}[ht]
  \centering
  \includegraphics[width=\columnwidth]{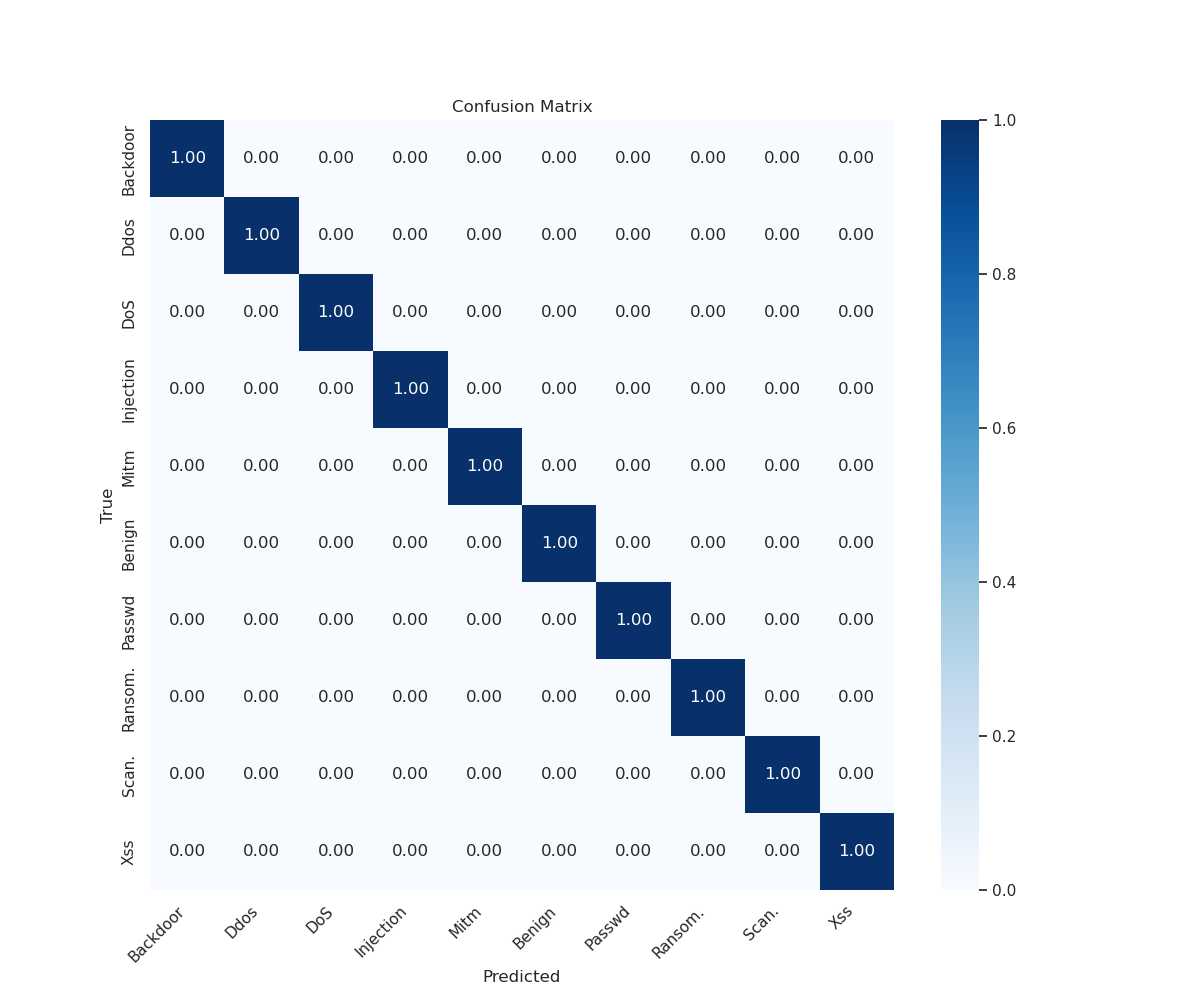}
  \caption{TON\textunderscore IoT: Confusion Matrix}
  \label{fig:tonfinegrained}
\end{figure}

Once we established that the algorithm can distinguish benign from malicious traffic with almost perfect accuracy, further examination was undertaken by training it with separate datasets designed specifically to each category of attack (Figure \ref{fig:tonfinegrained}). Our findings can be seen graphically herein.

At the onset of training a model, an impressive phenomenon emerges: its accuracy reaches perfection. This remarkable result can be attributed to generous sample sizes that permit nearly equitable undersampling across classes - an approach known as underbalancing which prevents certain underrepresented classes from exclusively appearing in training sets; post-balancing, test accuracy serves as an indicator of whether a model can generalize effectively across its full spectrum of potential output classes.

Cross-validation was employed as an additional measure to assess the quality of our model. Achieve perfect accuracy is rare in machine learning, prompting us to conduct this experiment one hundred times with unique random splits between training and test sets in each iteration, to ensure comprehensive test coverage. After calculation of average results, accuracy for our model was found to be an impressive 0.999936!

Although the classification accuracy may appear exceptional, it's essential to keep in mind that its authors achieved similar results using their best model. Furthermore, this dataset has been meticulously prepared for machine learning purposes and undergone special parsing processes as input to Threat Trekker algorithm; thus emulating the actual parsing process that will ultimately occur through connector middleware software. 

However, this result marks an impressive success and proves that with enough sample availability in place under favorable circumstances, full accuracy in classifying log and network data classification can indeed be reached.

%% file: sections/go_to_production.tex
As indicated earlier in this research paper, Threat Trekker was designed with one primary objective in mind - real-time identification and analysis of network anomalies and vulnerabilities. To accomplish this task, this algorithm utilizes log data accumulated from various applications or traffic captures which is then parsed through an API connector for analysis before becoming input for supporting model capabilities relevant for network surveillance/monitoring activities.

Please be mindful that this section enumerates a production vision for our algorithm, though no enhancements have yet been implemented at this initial stage. However, it must be acknowledged that its present iteration (i.e: managing data from multiple datasets), has proven its ability to successfully use machine learning algorithms to detect network hazards; consequently creating market opportunity.

\subsection{Model Training Methodology}\label{subsec:prod1}

Information is retrieved from network activities to the algorithm where it is used to train a machine learning model in incremental batches, continually refining accuracy and performance with time. By adapting to technology upgrades in response to network behavior change, an incremental training approach helps refine accuracy and performance over time while continuously learning from iteration within networks to detect abnormalities or vulnerabilities more efficiently. With dynamic learning process built-in, Threat Trekker always remains responsive to new network challenges while remaining effective against ongoing ones.

Precisely, this approach is especially well suited for distributed networks where logs are regularly created for monitoring purposes. Utilizing existing logs as training data for Threat Trekker allows seamless integration and ongoing optimization.

\subsection{Strategic Deployment Considerations for Business Logic}\label{subsec:prod2}

Thread hunting applications must handle an enormous workload by gathering data from all nodes within their network, but it must also take into account that such processing occurs simultaneously with other applications that generate network traffic, so their hosting platform should ideally feature some degree of flexibility to accommodate for changing resource needs effectively.

Cloud environments play an invaluable role here; not only are they conducive to developing asynchronous apps but they provide an ideal ecosystem for reactive ones as well \cite{reactive-manifesto}.
Threat Trekker inside of a container provides the optimal solution in this instance, taking full advantage of all its inherent strengths.
Containers provide seamless scalability for organizations, enabling them to easily adapt resource allocation in line with individual organizational needs and requirements. Their adaptability enables each business to determine how many resources it wishes to commit towards an algorithm's training; with batch training approaches able to intelligently drop random batches as the system nears its maximum resource threshold for maximum cost-efficiency and efficient utilization of computational capabilities.

Furthermore, deployment of connectors that facilitate data parsing from various applications into an algorithm presents an appealing option. Consolidating all necessary connectors within one cloud \textit{namespace} offers isolation of application logic, creating a more streamlined and cohesive cloud environment while simplifying management and maintenance duties while protecting each application independently from one another. Consequently, this cloud namespace serves as the central hub hosting Threat Trekker algorithm and its related connectors, facilitating efficient data exchange while increasing overall organizational scalability of this solution.

\subsection{Data Transmission and Integration with Threat Trekker}\label{subsec:prod4} 

Threat Trekker uses parquet files seamlessly, an intentional design choice made to enable compatibility with multiple connectors. Unfortunately, their development does not follow an ideal solution but must instead adapt based on specific network environments where they will be deployed - providing network administrators with the chance to leverage existing logic that they already own and reuse it as much as possible.

One particularly intriguing approach involves employing a distributed log system as the means of seamlessly transmitting logs generated from client applications to their respective connectors and filtering, parsing, aggregation, generating parquet files - ultimately serving as the foundational input needed for testing of models real time.

Connectors also possess the inherent capability of intaking network metrics stored as PCAP files, where these undergo extensive processing in order to extract and enrich log content for richer and deeper data analysis. Incorporating both network-level metrics as well as application log data provides a solid and holistic foundation upon which to build machine learning models for training or refinement purposes.

This approach requires creating an efficient and long-lasting mechanism to facilitate the transmission of large numbers of messages between networks, connectors, and the machine learning algorithm. At this juncture, Apache Kafka stands as an intelligent decision \cite{kafka-docs}. Widely considered a distributed streaming platform, Kafka also excels at log aggregation and management tasks. Provides an amazingly scalable and resilient infrastructure designed to collect, warehouse and archive logs in an error-tolerant fashion. This reservoir of stored log data serves as a consumable resource across applications; crucial in providing comprehensive analysis and insights development. 

Important components of this approach include creating Kafka producers within client network architecture to enable raw data transmission to designated connectors. While such activity requires proactive involvement from clients as it assumes responsibility for orchestrating data flow, such involvement also confers on them an unprecedented degree of control over data dissemination process - giving them more freedom than ever to implement appropriate masking/obfuscation mechanisms as necessary.

Granular control proves indispensable in situations in which network-generated information contains PII, subject to stringent data protection regulations. By providing masking processes prior to its usage by third-party algorithms, this approach aligns closely with privacy preservation imperatives mandated by existing data protection laws - making a delicate balance between proactive client engagement and safeguarding sensitive data an integral aspect of this architectural paradigm.

\begin{figure*}[ht]
  \centering
  \includegraphics[width=0.8\textwidth]{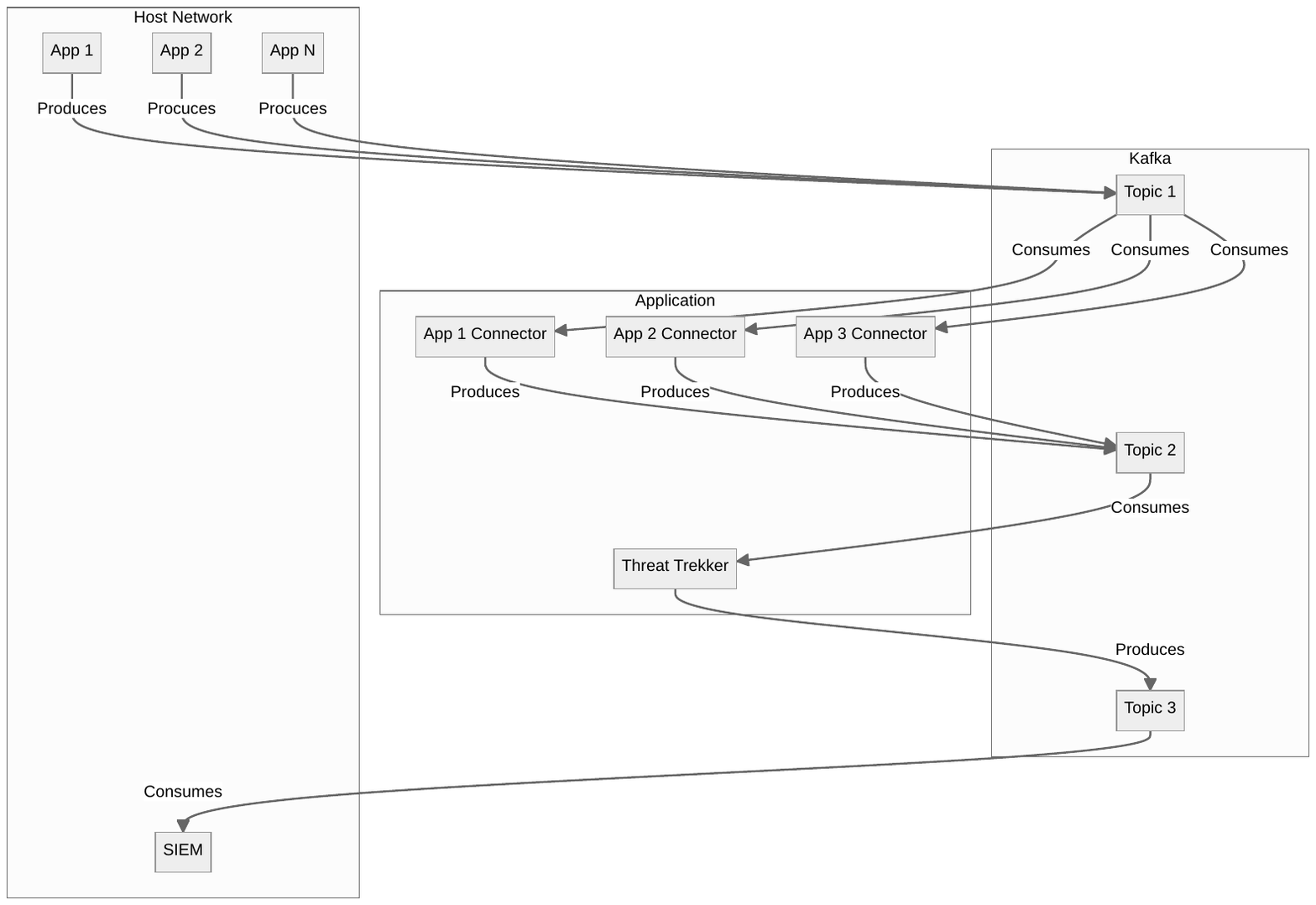}
  \caption{Overview of the Application Architecture}
  \label{fig:arch}
\end{figure*}

Figure \ref{fig:arch} provides an in-depth representation of the application architecture. It shows all of its interconnections among various nodes as well as Kafka's function of acting as the communication channel between nodes.

\subsection{Rationale for Implementing the Threat Trekker Algorithm: Justifying the Value Proposition}\label{subsec:prod3}

Utilizing distributed network logs, threat hunting algorithms, and SIEM technology simultaneously strengthens network security while taking an adaptive and pro-active approach to safeguard against potential vulnerabilities and threats.
Combining threat hunting algorithms with SIEM solutions makes for comprehensive network security monitoring. While threat hunting algorithms specialize in real-time anomaly detection and behavior analysis, SIEM allows for centralized log management, correlation and visualization of security events.

Threat hunting algorithms play an integral role in providing SIEM administrators with clear and actionable insights. By sending detailed and concrete data directly into a SIEM, an algorithm helps reduce noise while simultaneously drawing attention to critical security incidents - increasing its ability to respond rapidly and make important decisions more efficiently. Exploiting the Threat Trekker algorithm offers several distinct benefits, as it has the power to gain precise insight into users' behaviors and network usage patterns. By constantly learning and adapting to network dynamics, the algorithm gains an in-depth knowledge of legitimate user activities that allows it to make precise predictions as well as detect anomalies, intrusions or any instances of misuse effectively.

The algorithm's ability to differentiate normal from abnormal network behavior enables it to identify a variety of security threats, from known attacks such as DDoS attacks to novel, emerging attacks that appear overnight - contributing greatly to overall network security improvement. Real-time analysis ensures timely identification and response to new or emergent threats as they emerge, contributing significantly to network protection overall.

By taking advantage of this algorithm's sharp predictions, network administrators can proactively address potential security breaches and minimize risks quickly. Furthermore, its continuous learning and its ability to be upgraded over time makes it an indispensable component for strengthening network security measures.

Threat Trekker offers many distinct advantages to any organization that utilizes it, the chief being its ability to effectively understand user and network activity resulting in sharp predictions and accurate detection of intrusions or potential misuse, ultimately strengthening security posture while increasing resilience against various security threats.

%% file: sections/conclusion.tex
This paper proposes an innovative solution in threat hunting: an on-premise deployable application in real time. Experiments demonstrate how skewed attack distribution within datasets requires balancing to avoid overfitting during machine learning processing.

Threat Trekker could face potential challenges to its future prospects due to this factor. Eventually, as soon as it goes live in production mode, its metrics will gradually accumulate over time and accumulate enough samples so as to enable highly accurate predictions.
Prediction systems differ from traditional prevention systems by their ability to recognize subtle yet dangerous threats through anomalous behaviors detected within data flow or log information, providing network administrators with valuable time - and cost-saving options that counter potential vulnerabilities within their infrastructures. A proactive approach, this proactive strategy reduces downtime as well as cost.

Additionally, our results illustrate that, through careful data processing and application of learning algorithm techniques, remarkable predictive prowess in recognizing infrequent network attacks was demonstrated through these findings. Rare attacks with higher risks or relationships to APTs often prove as key targets of successful identification efforts.

An adverse consequence of using machine learning algorithms for classification may arise if they attempt to recognize hazards not included in their training dataset, creating an additional source of error and creating significant challenges when they attempt to categorize such behavior as anomalous. Although such behavior might be identified successfully as anomalous, classification might fail in categorizing it within one of the predefined output classes.

Overall, this approach seamlessly combines threat hunting with network framework by setting up an event stream channel for efficient data flow, as well as using machine learning models to classify potential hazards entailing further classification of possible threats and events. Furthermore, SIEM systems now can integrate and leverage this prediction algorithm's predictions directly.

%% file: sections/acknowledgments.tex
To my teacher, Alfonso for introducing me in the Threat Hunting World.